\documentclass{article}
\usepackage{amsfonts}
\usepackage{amssymb}
\usepackage{epsfig}
\newcommand{\ii}{{\rm i}}
\newcommand{\dx}{{\rm d}x}
\newcommand{\dy}{{\rm d}y}
\newcommand{\dt}{{\rm d}t}
\newcommand{\binom}[2]{\left(\!\!\begin{array}{cc}#1\\#2\end{array}\!\!\right)}
\newcommand{\piv}{\mbox{\boldmath$\pi$}}

\begin{document}
\title{Magnetic translations for a spatially periodic magnetic field}
\author{Wojciech Florek}
\date{Computational Physics Division, Institute of Physics, 
A. Mickiewicz University\\ ul.\ Umultowska 85, 61--614 Pozna\'n, Poland\\
{\it e-mail: florek@amu.edu.pl}}
\maketitle

\begin{abstract}
It is shown that in the case of free electron in a spatially periodic
magnetic field the concept of magnetic translations operators is still
valid and, moreover, these operators can be defined in the same way as
for a Bloch electron in a uniform magnetic field. The results can be
a useful tool in investigation of lately observed phenomena in 2DEG with
spatially modulated density.
\end{abstract}
  

\section{Introduction}
Recently there is a big interest in properties of charged particles in 
different inhomogeneous magnetic field profiles, what has been studied
both experimentally \cite{exp} and theoretically \cite{coop,theor}. 
Properties of the strongly correlated phases can be obtained from the 
photoluminescence spectra \cite{coop}, so much efforts are devoted to this 
technique \cite{PL}. This 
problem is strongly related to Anderson localization of electrons and, 
therefore, is important in investigations of high-$T_C$ superconductors and 
composite-fermions in the quantum Hall effect. 
Some recent experiments have shown that variation of density in 
two-dimensional electron systems lead to a fictitious {\em periodic}\/
magnetic field \cite{sept99} and a geometric resonance of the classical 
cyclotron orbit and the field period. A theory of this effect has been lately
proposed Zimbovskaya and Birman \cite{ZiBi}.

The aim of this paper is to consider a possibility of introducing magnetic 
translation operators in the case of a spatially periodic magnetic field and, 
if this introduction is possible, to determine their form. Some general
considerations on two-dimensional quantum systems in a singular vector 
potential {\bf A} were presented by Arai \cite{arai}, who generalized some 
concepts 
to {\em nonuniform magnetic fields} for continuous quantum systems and 
presented a method of reduction to lattice quantum systems. A special class
of introduced operators $T^{\bf A}$ leave invariant the Hilbert space
$L^2({\cal R}^2)$ (or $l^2({\cal Z}^2)$ for lattice systems) invariant, so
it is possible to construct a continuous version of a Hamiltonian of the 
Hofstader type \cite{kreft}. The problem analyzed in the present work is
simpler, however it is directly related with the above mentioned
phenomena. It is shown that magnetic translation operators introduced 
independently by Fishbeck \cite{fish}, Brown \cite{brow}, and Zak \cite{zaka}
for an electron in a uniform magnetic field and a periodic potential can
be also applied in the case of a spatially periodic magnetic field. In the 
simplest case a free electron is considered. 

In the next section a brief summary of the main results for a uniform
magnetic field is given, whereas the problem is solved in Sec.~3. Some 
remarks are presented in the last section.
 
\section{Bloch electron in homogeneous magnetic field}

For the sake of simplicity it is assumed that electron can move in
the two-dimensional plane $xy$, whereas the magnetic field 
${\bf H}=H\hat{{\bf z}}$ is perpendicular to it. Therefore, the
components $A_x$ and $A_y$ of a vector potential ${\bf A}$ are only 
relevant. In both cases, {\it i.e.}\ for homogeneous and inhomogeneous
magnetic field, the Hamiltonian is given by the well-known formula
 \begin{equation}\label{ham}
      {\cal H}={1\over{2m}}({\bf p}-{e\over c}{\bf A})^2+V({\bf r})\,.
 \end{equation}
 The only requirement is that $V({\bf r})$ is a periodic function of 
${\bf r}\in {\cal R}^2$. Positions of crystal nodes are determined
by vectors ${\bf R}$ of the two-dimensional translation group ${\cal T}$ 
isomorphic to ${\cal Z}\times{\cal Z}$, so they will be often replaced by 
pairs $(n_x,n_y)$ of integers, {\it i.e.}\ by their coordinates in the
crystal base $\{ {\bf a}_1,{\bf a}_2\}$. Since a point group symmetry
is not taken into account, the square lattice with ${\bf a}_1\perp{\bf a}_2$
and $a_1=a_2=a$ is considered. 

\paragraph {A form of the vector potential}
  
 For a constant and uniform magnetic field the vector potential ${\bf A}$ 
can be written as linear function of the position vector ${\bf r}$ 
\cite{floxx}. Moreover,
the coordinates $A_x$ and $A_y$ can be chosen in such a way that they do not 
depend on $x$ and $y$, respectively.\footnote{This condition is stronger than 
the radiation gauge $\nabla\cdot{\bf A}=0$.}
So in the most general case one obtains
 \begin{equation}\label{vecpo}
    A_x=\alpha y\qquad \mbox{and}\qquad A_y=\beta x\,.
 \end{equation}
  Real numbers $\alpha,\beta$ have to satisfy
 $$
   H=\beta-\alpha
 $$
 for a given magnitude $H$ of the magnetic field. Of course, this form 
includes the antisymmetric gauge $({\bf r}\times{\bf H})/2$ for $\alpha=
-\beta=-H/2$ and the Landau gauge for $\beta=H, \alpha=0$.

\paragraph{The symmetry of a problem}

  The periodicity of $V({\bf r})$ yields the two-dimensional translation
group ${\cal T}\simeq {\cal Z}\times {\cal Z}$ to be the symmetry 
group of the Hamiltonian. Therefore, one can introduce concepts of
quantized quasi-momentum, energy bands and Bloch electrons. Irreducible
representations of ${\cal T}$ label energy levels of ${\cal H}$. However,
in the presence of the magnetic field $H\hat{\bf z}$, at least except for
special values of $H$, projective representations have to be used and
their factor systems depend on the magnetic flux through the unit crystal 
cell. The periodic boundary conditions imposed on (projective) representation
$T({\bf R})$ leads to magnetic flux quantization and the concept of magnetic
cells (or magnetic periodicity) \cite{brow,floxx}. The periodic boundary 
conditions are also responsible for choosing the magnetic field perpendicular 
to a crystal plane \cite{brow,zaka,floacta2}.
It should be underlined that in Zak's approach projective representations of 
${\cal T}$ are replaced by vector representations of a central extension
of ${\cal T}$ by a group of factors included in U(1). Of course, these
two approaches are equivalent. 

\paragraph{Projective representations}
  
   Operators of a projective irreducible representation of ${\cal T}$
which commute with ${\cal H}$ can be chosen as \cite{fish,brow,zaka,floxx,BH}
 \begin{equation}\label{mtgop}
      T({\bf R}) = \exp[{\bf R}\cdot({\bf p}-{e\over c}{\bf A}')]\,,
 \end{equation}
 where ${\bf A}'$ is a vector potential associated with ${\bf A}$ and,
in this simplified considerations, given as
 \begin{equation}\label{assvecpo}
    A'_x=\beta y\qquad \mbox{and}\qquad A'_y=\alpha x\,.
 \end{equation}
  The commutation $T({\bf R})$ with ${\cal H}$ follows from the 
fact that coordinates of $\piv={\bf p}-{e\over c}{\bf A}$ commute with
coordinates of $\piv'={\bf p}-{e\over c}{\bf A}'$ \cite{brow,zaka,floxx}. 
Moreover,
the commutators $[\pi_x,\pi_y]$, $[\pi'_x,\pi'_x]$ {\it etc.} are numbers,
so they commute with any other operator. This fact is very important
in the derivation of a factor system for the representation $T({\bf R})$.
This factor system and the group-theoretical commutator 
$T({\bf R})T({\bf R}')T({\bf R})^{-1}T({\bf R}')^{-1}$ depend only on the
magnitude $H$, not on a form of the vector potential. However, working with
a local gauge ${\bf A}'_{\bf R}$ one can change a factor system, but the
commutator is unaffected (is gauge-independent) \cite{floacta}.

\paragraph {Movement of a Bloch electron}

   In the case of homogeneous magnetic field an electron moves around
the cyclotron orbit with coordinates of center given by operators 
$\pi'_y,\pi'_x$. This movement is quantized and is related to 
the broadening of the Landau levels \cite{brow,zakb}.

\section{Inhomogeneous magnetic filed}
 
   In order to keep symmetry described the translation group one has to
assume that non-constant terms in the magnetic field magnitude are periodic
with respect to $x$ and $y$ coordinates. Therefore, $\piv^2$, the generalized 
kinetic term of the Hamiltonian (\ref{ham}), is invariant under ${\bf R}\in 
{\cal T}$ and, at least in the first-order approximation, the potential
$V({\bf r})$ can be omitted. Applying 
the Fourier transform a periodic magnetic field can be written as a sum the
sine and cosine functions $\sin k_xx,\sin k_yy$ (or $\exp(\ii {\bf k}\cdot
{\bf r})$, if one prefers the complex analysis), so, to simplify a problem,
it is assumed that 
  \begin{equation}\label{inhomo}
        H=H_0+H_1\left[\cos(kx)+\cos(ky)\right]\,,
  \end{equation} 
 where $k=k_x=k_y=2\pi/a$.

\subsection{The vector potential}

For the magnetic field given by (\ref{inhomo}) the vector potential ${\bf A}$
can be chosen in many gauge invariant forms. For the sake of clarity 
the simplest form in the radiation gauge ($\nabla\cdot {\bf A}=0$) is assumed
 \begin{eqnarray}
     A_x&=&\alpha y - {H_1\over k} \sin(ky)\,,\nonumber \\
       \label{inpot}
     A_y&=&\beta x + {H_1\over k} \sin(kx)\,, 
 \end{eqnarray}
  with $\beta-\alpha=H_0$. 

\subsection {The symmetry group}

 Due to the periodicity of $H$ the symmetry group is still ${\cal T}\simeq
{\cal Z}\times{\cal Z}$. Its irreducible representations should label
eigensapces of ${\cal H}$. There is, however, one problem: {\em `Does it 
suffice
to consider projective representations or, due to higher terms of $x$ and $y$ 
in $A_x, A_y,$ and $H$, some more complex structure should be used?'}
Within the frame of Zak's approach it may mean that one has to
investigate non-Abelian extensions, for example. It seems that when for 
${\bf A}$ being linear function of ${\bf r}$ the second cohomology group
(related with projective representations or, equivalently, with central
extensions) comes into play, then for $A_x,A_y$ being second order functions 
of $x,y$ the third cohomology group should be considered {\it etc.} There are
some hints from mathematics and physics which indicate that it suffices to
limit to projective representations ({\it i.e.}\ the second cohomology 
group). 

At first, we are interested in ${\bf A}$ and $H$ expressed by the (co-)sine   
or other periodic functions with an infinite Taylor expansion, so---if the 
order of a cohomology group depends of the order of functions---cohomology
group of the infinite order should be taken into account. Moreover, factor
systems of projective representations in the case of a homogeneous magnetic
field depend on the magnetic flux through the unit crystal cell. This 
quantity is always a number calculated as an integral over the unit
cell of the product $H(x,y)\dx\dy$ (or as an integral over the edges of a cell
$\int\!{\bf A}({\bf r})\cdot {\rm d}{\bf r}$ \cite{arai}). Such an integral 
depends only on the 
constant term $H_0$ since the integration of periodic terms\footnote{In this 
place a {\em `periodic term'} denotes all but a constant term of the Fourier 
transform.} gives zero. The 
cohomology group of order $n$ demands considerations
of the $n$th co-boundaries and co-cycles, which involve $n$ lattice vectors. 
It seems it would be 
necessary in the case of a hypothetical `field' being a tensor of rank $n$ 
($F_{xyz}, G_{xyzt}$, {\it etc.})---in the case of magnetic field, which
can be described by the tensor $H_{xy}$, the most important are loops 
(drawn in two dimensions) encircling the magnetic flux (a one-dimensional 
object). At last, there are no well-investigated group-theoretical 
generalizations
of non-Abelian extensions corresponding to fourth, fifth {\it etc.} cohomology
groups (at least the author is not aware of such considerations).

\subsection{Projective representations}\label{prorep}

  In this section a form projective representation $T$ of ${\cal T}$ is 
derived. This representation should commute with ${\cal H}$ and
the results for the uniform magnetic field have to be revealed
in the limit $H_1\to 0$.

 Let us assume that operators of a projective representation have the
following form
 \begin{equation}\label{newmtg}
 T({\bf R})=\exp\left(-{\ii\over\hbar}\piv'\cdot{\bf R}\right)\,,
 \end{equation}
 where a form the operator  $\piv'$ depends on ${\bf A}$ and in the limit 
$H_1\rightarrow 0$ we have $\piv'=p-(e/c){\bf A}'$ with ${\bf A}'$ defined
in the previous section by Eq.\ (\ref{assvecpo}). Due to the periodicity
of $V({\bf r})$ (moreover, it is assumed $V({\bf r})=0$ in the simplest
approximation) it is enough to calculate a commutator $[T({\bf R}),\piv^2]$.
  If $\pi_x$ and $\pi_y$ commute with $T({\bf R})$ then also ${\cal H}$ does. 
The condition $[T({\bf R}),\pi_x]=[T({\bf R}),\pi_y]=0$ allows also labeling 
of eigenspaces of the canonical momenta $\pi_x$ and $\pi_y$ by irreducible 
projective representations of ${\cal T}$. Substituting  
 $
   P=-{\ii\over\hbar}\piv'\cdot{\bf R}
 $
 one can write  
 \[ T({\bf R})=\exp P = \sum_{n=0}^\infty {{P^n}\over{n!}}
 \]
 and 
 \begin{equation}\label{Pcom}
 [\exp P,\pi_\xi]=\sum_{n=0}^\infty {1\over{n!}}[P^n,\pi_\xi]\,,\qquad
  \xi=x,y\,.
 \end{equation}
 The last commutator equals
 \begin{equation}\label{Pncom}
  [P^n,\pi_\xi]=
 \sum_{k=0}^{n-1} \binom{n}{k}C_{n-k}^\xi P^k\,,
 \end{equation}
 where
 \[
  C_l^\xi=[P,C_{l-1}^\xi]\qquad \mbox{and} \qquad 
       C_0^\xi=\pi_\xi\,.
 \]
   These formulae yield that $C_1^\xi=[P,\pi_\xi]=0$ is a sufficient 
condition
for $[T({\bf R}),{\cal H}]=0$. For example, it can be solved in the case of 
the homogeneous magnetic field, {\it i.e.}\ when $\piv$ is a linear function
of $x$ and $y$, and this solution was found by Brown and Zak and then 
generalized to any linear vector potential by the author (see also 
\cite{fish}). However, the
case of the inhomogeneous magnetic field yields ${\bf A}$ being a square
(or higher order) function of coordinates $x,y$ and the condition 
$C_1^x=0$ does not lead to any non-trivial solutions for $P$. Therefore,
the formula (\ref{Pncom}) has to be substituted to the condition (\ref{Pcom})
and the solution has to be found in this more general case. So, one obtains
 \begin{equation}\label{Pnsum}
 [\exp P,\pi_\xi]=\sum_{n=0}^\infty {1\over{n!}} 
       \sum_{k=0}^{n-1} \binom{n}{k}C_{n-k}^\xi P^k
 = \left(\sum_{n=1}^{\infty} {1\over{n!}} C_n^\xi\right) \exp P \,.
 \end{equation}
  Therefore, the sum of commutators $C_n^\xi$ has to be equal to zero, what,
is satisfied by the special solution $C_1^\xi=0$.
After substitution $\pi'_\xi =-\ii\hbar\partial_\xi-(e/c)A'_\xi$ 
 the operator $P=-{\ii\over\hbar}X\pi'_x -{\ii\over\hbar}Y\pi'_y$,
 where $X=n_xa$, $Y=n_ya$ are coordinates of ${\bf R}$, 
decomposes into two summands:
 \begin{equation}\label{Ppart}
  P=P_1+P_2\,;\qquad
P_1=-(X\partial_x+Y\partial_y)\,,\quad 
P_2={{\ii e}\over{\hbar c}}(XA'_x +YA'_y)\,.
 \end{equation}
 So, in a general case, we have
 \begin{equation}\label{compp}
  C_1^\xi=[P,\pi_\xi]=
 {e\over c}\left[ (X\partial_x +Y\partial_y)A_\xi 
  - \partial_\xi (XA'_x +YA'_y)\right]\,.
 \end{equation}
 If one assumes that $\partial_x A_x=\partial_y A_y=0$, this can be 
simplified to 
 \begin{eqnarray}
  C_1^x&=& {e\over c}[Y\partial_yA_x-\partial_x (XA'_x+YA'_y)]\,;\nonumber \\
  C_1^y&=& {e\over c}[X\partial_xA_y-\partial_y (XA'_x+YA'_y)]\,.
 \label{compps}
 \end{eqnarray}
\smallskip

\paragraph{Example} Let us consider the case $A_x=\alpha y$ and $A_y=\beta x
+H_1 x^2/2$. We are looking for a solution in a similar form, {\it i.e.}\
determined by an associated vector potential ${\bf A}'$ being a square function
of $x$ and $y$, so we assume that 
 \[
  A'_x=\alpha' y + \alpha'' y^2\qquad \mbox{and} \qquad
  A'_y=\beta' x + \beta'' x^2\,.
\]
 The equation (\ref{compps}) gives
 \begin{eqnarray*}
  C_1^x&=& {e\over c}Y(\alpha-\beta'-2\beta''x)\,;\\
  C_1^y&=& {e\over c}X(\beta+H_1x-\alpha'-2\alpha''y)\,.
 \end{eqnarray*}
 Since $C_1^\xi\neq 0$ in a general case, we have to calculate $C_2^\xi$
(note that only $P_1$ is relevant):
 \begin{eqnarray*}
  C_2^x&=& [P,C_1^x]=2\beta''{e\over c}XY\,;\\
  C_2^y&=& [P,C_1^y]={e\over c}X(-XH_1+2Y\alpha'')\,.
 \end{eqnarray*}
  Therefore $C_l^\xi=0$ for $l\ge3$ and the commutator $[\exp P,\pi_\xi]=0$
if $C_1^\xi+C_2^\xi/2=0$ for any $x,y,X,Y$. For $\xi=x$ we obtain 
 $$ \beta''=0\qquad\mbox{and}\qquad\beta'=\alpha\,. $$
 It is easy to notice that a solution for $\xi=y$ and $H_1\neq0$ can be 
obtained only when $\alpha''=0$ and $X=0$. It means that (magnetic) 
translations $T([0,Y])$ commute with the Hamiltonian. It is not surprising 
since the magnetic field $H=(\beta-\alpha)+H_1x$ is not periodic in the $x$-th
direction.
\bigskip

This example shows that the linear term in $A_\xi$ (the constant term in $H$) 
always appear in $A'_\xi$ in the same way as in the case of the  
homogeneous magnetic field. Therefore, for the vector potential given by
Eq.~(\ref{inpot}) we choose the associated potential in the following form
 \begin{eqnarray}
     A'_x&=&\beta y + f(x,y)\,,\nonumber \\
       \label{asinpot}
     A'_y&=&\alpha x + g(x,y)\,, 
 \end{eqnarray}
 where functions $f$ and $g$ will be determined from the commutation 
conditions. Substituting it to Eq.~(\ref{compps}) we obtain
 \begin{eqnarray}
  C_1^x&=& -{e\over c}\left\{YH_1\cos(ky)
      +\partial_x [Xf(x,y)+Yg(x,y)]\right\}\,;\nonumber \\
  C_1^y&=& \phantom{-}{e\over c}\left\{XH_1\cos(kx)
      -\partial_y [Xf(x,y)+Yg(x,y)]\right\}\,.
 \label{rompps}
 \end{eqnarray}
  The next commutators will determined with the use of the operator $P_1$,
see Eq.~(\ref{Ppart}), and derivatives of each function will be consider 
separately. The first part of $C_1^x$, up to the constant factor 
$-H_1e/kc$, gives the following series
 $$
  kY\cos(ky)\,,\;\;
  (kY)^2\sin(ky)\,,\;\;
  -(kY)^3\cos(ky)\,,\;\;
  -(kY)^4\sin(ky)\,,\;\;
  (kY)^5\cos(ky)\,,\,\dots
 $$
 Decomposing it to two series with $\cos(ky)$ and $\sin(ky)$, respectively,
and substituting it to the infinite sum in Eq.~(\ref{Pnsum}) one obtains
(taking into account that $k=2\pi/a$ and $Y=n_ya$, so $kY=2n_y\pi$)
 \begin{eqnarray}
&&\cos(ky) \sum_{n=0}^{\infty} 
{ {(-1)^n(kY)^{2n+1}} \over{(2n+1)!}} 
+\sin(ky) \left(1-\sum_{n=0}^{\infty} 
{ {(-1)^n(kY)^{2n}} \over{(2n)!}}\right)\nonumber\\[3pt]
\label{sinky}
&=& \cos(ky)\sin(kY)+\sin(ky)\big(1-\cos(kY)\big)=0\,.
 \end{eqnarray}
 The same result is obtained for a part of $C_1^y$ containing $\cos(kx)$.
 Therefore, operators $T({\bf R})$ commute with the Hamiltonian for 
{\em trivial}\/ functions 
  $$
   f(x,y)\equiv g(x,y)\equiv 0\,.
  $$
  It is not so surprising if we recall that the factor system depends on a 
magnetic flux through a lattice cell \cite{floacta} and the periodic part of 
$H$ gives no input to it. Such solution leads to interesting difference between
$H$ and $H'$: the first is periodic, whereas the 
second is uniform since $H'=-H_0$. This difference 
is more evident if we take into account the vector potentials ${\bf A}$ and 
${\bf A}'$: the original one satisfy an inhomogeneous wave equation 
\cite{jack}:
 $$
  \nabla^2{\bf A}=-{{4\pi}\over{c}}{\bf J}=kH_1[\sin(ky),-\sin[kx)]\,,
 $$ 
 whereas $\nabla^2{\bf A}'={\bf 0}$. 

The obtained solution means that a spatially periodic magnetic field does not
require any changes in the group of magnetic translations (projective 
representations of the translation group). 

\subsection{Other solutions}

It is interesting to check what happen in the case when $A'_\xi$ depends
on $x$ and $y$ in a similar way as $A_\xi$. To begin with 
we choose 
 \begin{equation}\label{fg}
   f(y)= {H_1\over k} \sin(ky)\,,\qquad
     g(x) = -{H_1\over k} \sin(kx)\,.  
 \end{equation}
  Such a choice leads to the associated magnetic field $H'$ equal to
 $$
   H'=(\alpha-\beta)-H_1[\cos(kx)+\cos(ky)]=-H\,,
 $$
 as we have obtained in the case of the homogeneous magnetic field ({\it cf.}\
\cite{floacta}). The series of commutators $C_l^x$ containing $g(x)$ and its
derivatives is calculated as follows (up to the constant factor $H_1Ye/kXc$)
 $$
  kX\cos(kx)\,,\;\;
  (kX)^2\sin(kx)\,,\;\;
  -(kX)^3\cos(kx)\,,\;\;
  -(kX)^4\sin(kx)\,,\;\;
   (kX)^5\cos(kx)\,,\;\;\dots
 $$
  Substituting it to the infinite sum in Eq.~(\ref{Pnsum}) one obtains, as
in Eq.\ (\ref{sinky}),
 $$
 \cos(ky)\sin(kY)+\sin(ky)\big(1-\cos(kY)\big)=0\,.
 $$
 However, this solution has been obtained after division by $X$, so it is valid
only for $X\neq0$. Analogous considerations for $f(y)$ lead to the same 
result, but now $Y\neq0$. Therefore, the potential obtained for functions
(\ref{fg}) determines operators commuting with the Hamiltonian only for
$X,Y\neq0$ or $X=Y=0$. Exclusion of the axis $X=0$ is caused by the lack of 
one power of $X$. It can be revealed if the magnetic translations will be 
defined {\em locally}\/ \cite{floacta}, {\em i.e.}\/ a form of functions $f$ 
and $g$ will
depend on $X$ and $Y$. The simplest solution is to put
 $$ f_Y(y)=Yf(y)\qquad\mbox{\rm and}\qquad g_X(x) =Xg(x)\,. $$
 Another way is realized by switching $f$ and $g$, so
 \begin{equation}\label{fgp}
   f(x)= {H_1\over k} \sin(kx)\,,\qquad
     g(y) = -{H_1\over k} \sin(ky)\,.  
 \end{equation}
 In this case, however, $H'=-H_0$ (without the periodic term) and, moreover
the vector potential ${\bf A}'$ is no longer written in the radiation gauge.

\subsection{Movement of a Bloch electron}

\begin{figure}
\begin{center}
\epsfxsize=7cm
\epsfbox{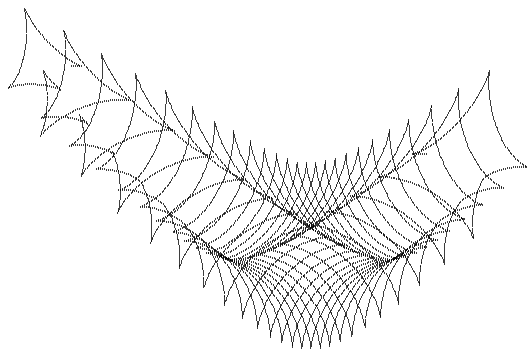}
\end{center}
\caption[]{A trajectory of the orbit center for a large orbit radius $\rho\gg
a$\label{fig1}}
\end{figure}

The approximated considerations of an electron in spatially inhomogeneous 
field presented in many textbooks, {\it e.g.} \cite{jack}, show that for small
values of $\nabla H$ a charged particle gains an additional velocity 
perpendicular to ${\bf H}$ and $\nabla H$. Since $\nabla H$ is perpendicular
to lines on which ${\bf H}$ is constant then it is most likely that the orbit
center will move along these lines. Results of simple numerical simulations 
(only the Lorentz force for a periodic magnetic field has been taken into 
accout ) seem to confirm this 
picture. However, a system is very unstable and actual behavior strongly
depends on starting values and accuracy. When a time-step has been relatively 
large ($\omega\Delta t\simeq0.01$), the center moved towards the maximum (for 
large $H$) or minimum (for small $H$) of the function $H(x,y)$. This movement
was almost negligible for small values of $\Delta t$ 
($\omega\Delta t\simeq0.0001$). A critical value of $H$ depends on a particle
velocity and corresponds to the orbit diameter $2\rho$ close to the field 
period, {\it cf.} \cite{sept99}.
It could be expected that in the limit case $\Delta t\to \dt$ the orbit center
would be `stabilized' on an `equipotential' line $H=${\it const}. The other 
conclusions are almost obvious: (i) in a strong magnetic field (a very small 
orbit diameter) the movement of the orbit center is much slower since
a particle moves in almost constant magnetic field; (ii) in the case of
small magnetic field (a large orbit diameter) a particle moves in regions
of different magnetic field magnitudes and the orbit center trajectory
is not a smooth line (see Fig.~\ref{fig1}).  

\section{Final remarks}
 It has been shown that a spatially periodic magnetic field leads to the
same magnetic translation operators as in the case of a uniform field and
a periodic potential $V({\bf r})$. Since the results obtained do not 
depend on an actual value of $k$ then they are valid for any periodic
function of $x$ and $y$ written as its Fourier transform. When a periodic
potential $V({\bf r})$ is also present than our considerations are applicable
to the case of commensurate periods of $H$ and $V$ only. The results of
numerical simulations suggest that the orbit center moves along an 
`equipotential' line $H=${\it const}. This movement is more stable in the
case of relatively large fields, {\it i.e.} for the orbit radius $\rho$ 
smaller than the field period $a$. The magnetic translation operators cannot
be interpreted as the position of the orbit center \cite{floacta,JL}, since 
the later is not a constant of motion. They rather correspond to the center 
of the orbit center trajectory. It should be underline that this work
does not present a theory of the observed effects \cite{sept99}, but 
provides a mathematical tool, which can be useful in such investiagtions.

\section*{Acknowledgments}
 This work is supported by the State Committee for Scientific 
Research (KBN) within the project No 263/P03/99/16.

\end{document}